\documentclass[12pt]{JHEP3}
\usepackage{latexsym,graphicx,epsfig,psfrag,here}

\newcommand{\beq}{\begin{equation}} 
\newcommand{\eeq}{\end{equation}} 
\newcommand{\beqa}{\begin{eqnarray}} 
\newcommand{\eeqa}{\end{eqnarray}} 
\newcommand{\beqan}{\begin{eqnarray*}} 
\newcommand{\eeqan}{\end{eqnarray*}} 
\newcommand{\ba}{\begin{array}} 
\newcommand{\ea}{\end{array}} 
\newcommand{\no}{\nonumber}

\newcommand{\ol}{\overline}

\newcommand{\ve}{\varepsilon}

\newcommand{\cL}{{\cal L}}

\newcommand{\mbf}{\mathbf} 
 
\newcommand{\dfrac}{\displaystyle \frac}

\newcommand{\nn}{\nonumber \\}

\newcommand{\bea}{\begin{eqnarray}} 
\newcommand{\eea}{\end{eqnarray}} 
 
\newcommand{\fsl}{\not\!}

\def\pmb#1{\setbox0=\hbox{#1}%  poor man's bold macro
    \kern-.025em\copy0\kern-\wd0
    \kern.05em\copy0\kern-\wd0
    \kern-.025em\raise.0433em\box0}

\newcommand{\PL}[3]{{Phys. Lett.} {\bf#1} {(#2)} {#3}} 
\newcommand{\PRL}[3]{{Phys. Rev. Lett.}  {\bf#1} {(#2)} {#3}} 
\newcommand{\PR}[3]{{Phys. Rev.} {\bf#1} {(#2)} {#3}} 
\newcommand{\NP}[3]{{Nucl. Phys.} {\bf#1} {(#2)} {#3}} 
\newcommand{\EPJ}[3]{{Eur. Phys. J.} {\bf#1} {(#2)} {#3}} 
\newcommand{\ZP}[3]{{Z. Phys.} {\bf#1} {(#2)} {#3}} 

\preprint{IFIC/02-32\\
UWThPh-2002-16\\ 
July 2002\\ }

\title{\Large \bf Radiative \pmb{$\tau$} ~Decay \\ and the 
Magnetic Moment of the Muon  
\thanks{Work supported in part by TMR, EC-Contract  
No. ERBFMRX-CT980169 (EURODA$\Phi$NE).} \\}

\author{ V. Cirigliano \\
Departament de F\'{\i}sica Te\`orica, IFIC, Universitat de 
Val\`encia - CSIC\\ 
Apt. Correus 2085, E-46071 Val\`encia, Spain \\
E-mail: \email{Vincenzo.Cirigliano@ific.uv.es} }

\author{G. Ecker  and H. Neufeld \\ 
Institut f\"ur Theoretische Physik, Universit\"at 
Wien\\ Boltzmanngasse 5, A-1090 Vienna, Austria \\
E-mail: \email{ecker@thp.univie.ac.at},  
\email{neufeld@thp.univie.ac.at}
}
 
\abstract{
We discuss the decay $\tau^- \to \nu_\tau \pi^- \pi^0 \gamma$ in terms
of a model with the correct low-energy structure and with the relevant
resonance degrees of freedom. The nontrivial radiative dynamics
becomes visible for large photon momenta only. We use the model to
calculate electromagnetic corrections for the two-pion contribution to
hadronic vacuum polarization extracted from photon-inclusive two-pion
decays. The corrections are insensitive to the details of the model
and depend on the pion form factor only.
Putting all relevant isospin violating corrections together,
we obtain a shift $\Delta a_\mu = (- 120 \pm 26 \pm 3)\times 10^{-11}$ 
to be applied to determinations of the anomalous magnetic moment of 
the muon from photon-inclusive $\tau$ data.}

\begin{document} 

\section{Introduction}
\label{sec:intro}
\renewcommand{\theequation}{\arabic{section}.\arabic{equation}}
\setcounter{equation}{0}

The process $\tau^- \to \nu_\tau \pi^- \pi^0 \gamma$ involves an
interplay of strong, weak and electromagnetic effects. 
For small momentum transfer to the hadronic system as in  
the related radiative pionic beta decay $\pi^+ \to e^+ \nu_e \pi^0 
\gamma$, the theoretical determination of the associated decay amplitude 
is well under control. 

At low energies, the standard model of strong and electroweak
interactions is described by an effective quantum field theory called
chiral perturbation theory (CHPT) \cite{gl8485a}. It treats the
interactions of the experimentally observed hadronic degrees of freedom 
at low energies by exploiting the constraints imposed by the
symmetries of the standard model, in particular the spontaneously 
broken chiral symmetry.  
The application of CHPT is not restricted to purely strong interaction 
processes. With appropriate external source terms,
also semileptonic and electromagnetic reactions involving leptons
can be investigated.

The applicability of CHPT ends once typical hadronic energies are 
approaching the mass of the $\rho$ resonance. Nevertheless, the
correct low-energy limit, unambiguously described by CHPT, puts  
constraints on the behaviour of amplitudes also in the resonance
region. The low-energy constants (LECs) of order $p^4$ in the chiral 
expansion are known to be dominated by the low-lying meson 
resonances \cite{egpr89}. This correspondence facilitates the matching 
between the low-energy domain and the resonance region. 
Applications along these lines comprise the pion form factor 
\cite{spanff1,spanff2}, radiative $\rho$ decays \cite{hn95} and four-pion 
production \cite{eu02}.

The radiative $\tau$ decay is a further process where the resonance 
regime of the standard model can be tested. A second motivation for 
the present investigation lies in the relevance of the 
non-radiative decay $\tau^- \to \nu_\tau \pi^- \pi^0$ for a precise
determination of hadronic vacuum polarization, which is essential for the 
calculation of both the energy dependence of the QED fine structure 
constant and of the anomalous magnetic moment of the muon.
The reason is that a CVC relation between electromagnetic and weak form 
factors allows to relate the differential decay rate of  
$\tau^- \to \nu_\tau \pi^- \pi^0$ to the cross section 
$\sigma(e^+ e^- \to \pi^+\pi^-)$ in the isospin limit. 

However, the present theoretical and experimental accuracy of hadronic
vacuum polarization requires a precise determination of all isospin 
breaking effects due to $m_u \ne m_d$  and electromagnetism. The
radiative corrections for $\tau^- \to \nu_\tau \pi^- \pi^0$ are part
of this task. To cancel the infrared divergences generated by 
the one-loop exchange of a virtual photon, the associated 
radiative decay mode must be included.  

In the low-energy limit, the necessary modifications of the CHPT
framework in the presence of virtual photons \cite{Urech,NeuRup} and 
virtual leptons \cite{KNRT00} have already been worked out.
In this paper we want to cover 
also the resonance region relevant for the two-pion decay of the $\tau$ 
lepton. Our present work is an extension of a previous letter \cite{cen1}
where electromagnetic corrections and their impact on the
determination of $a_\mu$ had already been discussed. However, the
analysis of Ref.~\cite{cen1} carried a rather large uncertainty
because of the incomplete matching between virtual and real
photons that depends on the experimental setup. In the present paper,
we discuss in detail the radiative corrections for the case where
all radiative events are included in the two-pion data sample. This is
a case of immediate practical interest because it corresponds to the
analysis of the ALEPH experiment \cite{ALEPH}\footnote{We are grateful 
to Michel Davier and Andreas H\"ocker for information on the ALEPH 
procedure.}.

The paper is organized as follows. In Sec.~\ref{sec:amp}  we discuss 
the general structure of the $\tau^- \to \nu_\tau \pi^- \pi^0 \gamma$ 
amplitude together with the dominant contributions in the resonance 
region. The decay rate is calculated in Sec.~\ref{sec:rates}. Applying 
different cuts in the photon energy, 
we explore the energy region where deviations from bremsstrahlung can 
be expected. In Sec.~\ref{sec:vacpol} we turn to the calculation of 
radiative corrections for the $\tau$ decay into two pions relevant for  
hadronic vacuum polarization. The isospin violating
corrections for $a_\mu$ are collected in Sec.~\ref{sec:amu}. Finally, 
our conclusions are summarized in Sec.~\ref{sec:conc}. The chiral
resonance Lagrangian, kinematical details and loop functions are
relegated to two appendices.

\section{Amplitude for \pmb{$\tau^- \to \nu_\tau \pi^- \pi^0 \gamma$}}
\label{sec:amp} 
\renewcommand{\theequation}{\arabic{section}.\arabic{equation}}
\setcounter{equation}{0}

The matrix element for the decay
$$
\tau^-(P) \to \nu_\tau(q) \pi^-(p_-) \pi^0(p_0) \gamma(k)
$$
has the general structure
\beqa
T & = & \left. e G_F V^*_{ud} \ve^{\mu}(k)^*\right\{
F_\nu \ol{u}(q) \gamma^{\nu} (1 - \gamma_5) 
(m_\tau + \fsl P - \fsl k) \gamma_{\mu} u(P) \label{eq:T}\\
&+&\left.  (V_{\mu\nu} - A_{\mu\nu}) \ol{u}(q) \gamma^{\nu} 
(1 - \gamma_5) u(P) \right\} \no ~.
\eeqa
The first part of this matrix element describes bremsstrahlung off
the initial $\tau$ lepton with
\begin{equation}
F_\nu = (p_0 - p_-)_\nu f_+(t)/2 P\cdot k~, \qquad t=(p_- + p_0)^2~.
\label{eq:initbrems}
\end{equation}
The form factor $f_+(t)$ governs the non-radiative decay. It is
identical to the electromagnetic form factor of the pion
because we work in the isospin limit $m_u=m_d$.

The second part of Eq.~(\ref{eq:T}) describes the vector and
axial-vector components of the transition
$$
W^-(P-q) \to \pi^-(p_-) \pi^0(p_0) \gamma(k) ~.
$$
The hadronic tensor $V_{\mu\nu}$ contains bremsstrahlung
off the $\pi^-$ in the final state. Gauge invariance implies the Ward 
identities
\beqa
k^\mu V_{\mu\nu} & = & (p_- - p_0)_\nu f_+(t) \label{eq:Ward} \\*
k^\mu A_{\mu\nu} & = & 0~.  \no
\eeqa

The tensor amplitudes $V_{\mu\nu}, A_{\mu\nu}$ contain four invariant
amplitudes each \cite{beg93}.  For $V_{\mu\nu}$ we use a rather 
special form here that is convenient for our purposes. In
$A_{\mu\nu}$ we only include the two amplitudes that actually arise in
our calculation:
\beqa
V_{\mu\nu} & = & f_+[(P-q)^2]\displaystyle\frac{p_{-\mu}}{p_-\cdot k}
(p_-+k-p_0)_\nu - f_+[(P-q)^2]g_{\mu\nu} \label{eq:vmunu} \\*
&+&  \displaystyle\frac{f_+[(P-q)^2]-f_+(t)}{(p_-+p_0)\cdot k} 
(p_-+p_0)_\mu(p_0-p_-)_\nu  \nn
&+&  v_1 (g_{\mu\nu}p_-\cdot k - p_{-\mu}k_\nu) +
v_2 (g_{\mu\nu}p_0\cdot k - p_{0\mu}k_\nu) \nn
&+&  v_3 (p_{-\mu}p_0\cdot k - p_{0\mu}p_-\cdot k)p_{-\nu}
+ v_4 (p_{-\mu}p_0\cdot k - p_{0\mu}p_-\cdot k)(p_- + p_0 + k)_\nu \nn
A_{\mu\nu} & = & i a_1 \ve_{\mu\nu\rho\sigma} (p_0 - p_-)^\rho
k^\sigma + i a_2 (P-q)_\nu \ve_{\mu\rho\sigma\tau}k^\rho
p_-^\sigma p_0^\tau \label{eq:amunu} ~.
\eeqa
These tensor amplitudes have the following properties:
\begin{itemize}  
\item $V_{\mu\nu}$ and  $A_{\mu\nu}$ satisfy the Ward identities
(\ref{eq:Ward}) for any $v_1, \dots, v_4, a_1, a_2$.
\item Taking into account $(P-q)^2=t+2(p_-+p_0)\cdot k$,
Low's theorem \cite{low} is manifestly satisfied:
\begin{eqnarray} 
V_{\mu\nu} &=& f_+(t) \frac{p_{-\mu}}{p_-\cdot k}(p_- - p_0)_\nu
\label{eq:Low} \\
& +& f_+(t) \left(\frac{p_{-\mu}k_\nu}{p_-\cdot k} - g_{\mu\nu}\right)
\nn
& + & 2 \frac{df_+(t)}{dt}\left(\frac{p_{-\mu}p_0\cdot k}{p_-\cdot k}
- p_{0\mu}\right)(p_- - p_0)_\nu + O(k) ~.\no
\end{eqnarray} 
The invariant amplitudes  $v_1, \dots, v_4$ as well as $a_1, a_2$ 
remain finite for $k\to 0$. Of course, initial state bremsstrahlung
in Eqs.~(\ref{eq:T},\ref{eq:initbrems}) contributes also in the Low
limit. 
\item The leading order in the low-energy expansion corresponds to
\begin{eqnarray}
 & f_+=1 \label{eq:p2} \\
& v_1=v_2=v_3=v_4=0 \nn
& a_1=a_2=0 ~.\no
\end{eqnarray}
\end{itemize} 

Our aim is to construct a model for radiative $\tau$ decay that can 
be trusted in all of phase space. A major constraint is the correct
low-energy limit to $O(p^4)$. The local contribution of $O(p^4)$
contains information on which meson resonances must be taken into
account for extending the chiral amplitude into the resonance region. 
On the other hand, except for the pion form factor $f_+(t)$, we are
not going to include all one-loop contributions arising at $O(p^4)$, 
let alone higher orders. Since we want to use the model for all energies
accessible in $\tau$ decay a complete one-loop calculation would not
be of much help anyway. Part of the loop contributions will be
contained in the energy-dependent widths of the resonance propagators 
and in the representation of the pion form factor of
Ref.~\cite{spanff1} that we shall use. Therefore, our approach uses the 
low-energy limit
to identify the relevant degrees of freedom but the model goes clearly
beyond the range of applicability of the low-energy expansion
of QCD. Such a procedure works very well for the pion form factor 
itself \cite{spanff1,spanff2} and it also seems to be successful in 
radiative $\rho$ decays \cite{hn95} and in four-pion 
production \cite{eu02}.

In addition to the one-loop amplitude, $V_{\mu\nu}$ depends to
$O(p^4)$ on the LECs $L_9, L_{10}$
\cite{gl8485a}:
\begin{eqnarray} 
f_+(t) &=& 1 + 2 L_9 t/F_\pi^2 \nn
v_1 &=& 4 (2 L_9 + L_{10})/F_\pi^2 \label{eq:p4} \\
v_2 &=& 4 L_{10}/F_\pi^2 \nn
v_3=v_4 &=& 0 ~. \no
\end{eqnarray}
The constants $L_9, L_{10}$ can be interpreted 
as low-energy remnants of $\rho$ and $a_1$ exchange \cite{egpr89}. 
As usual, $\rho$ exchange will turn out to dominate by far and the
$a_1$ contribution is in practice undetectable in this decay. 

For completeness, we include also the (leading) contributions of 
$O(p^4)$ to $A_{\mu\nu}$ due to the chiral anomaly. The diagrams in 
Fig.~\ref{fig:anom} get contributions from the Wess-Zumino-Witten 
functional \cite{WZW}:
\begin{eqnarray} 
a_1=\displaystyle\frac{1}{8\pi^2 F_\pi^2}~,\qquad
a_2=- \displaystyle\frac{1}{4\pi^2 F_\pi^2 [(P-q)^2-M_\pi^2]}~.
\end{eqnarray} 
By itself, the anomaly contribution cannot be expected to
yield a realistic axial amplitude for all available energies. We take
it rather as an indication of the expected order of 
magnitude of $A_{\mu\nu}$. Anticipating the result of the numerical
calculation, the anomalous amplitude is very much suppressed compared to 
the dominant $\rho$ exchange contributions and turns out to be 
completely negligible in radiative $\tau$ decay.
\FIGURE[ht]{
\epsfig{file=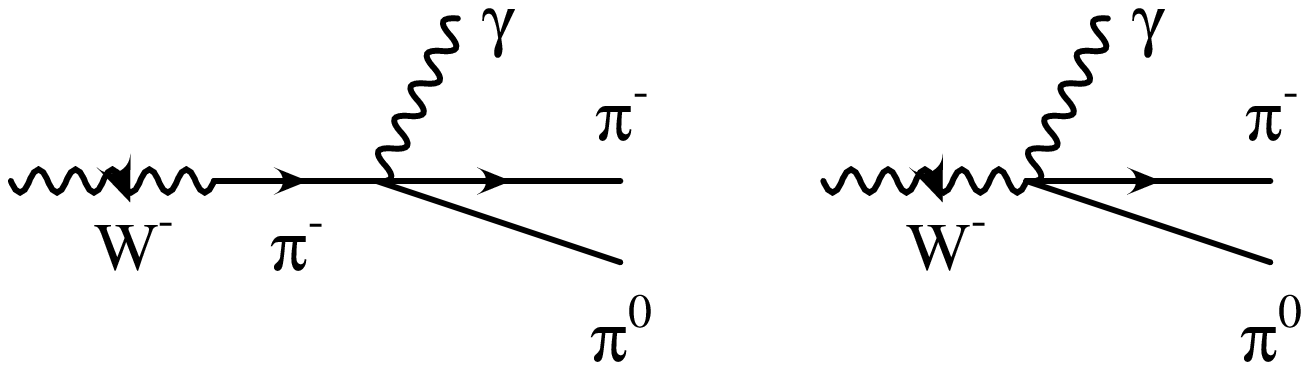,height=4cm}
\caption{Anomalous diagrams for the transition $W^- \to \pi^-\pi^0
\gamma$ contributing to the axial tensor amplitude $A_{\mu\nu}$.}
\label{fig:anom} 
}

The amplitude of $O(p^4)$ in (\ref{eq:p4}) tells us
that $\rho$ and $a_1$ exchange contribute to
the vector amplitude at low energies. We now assume that
these are the dominant degrees of freedom for all accessible
energies. As is easy to understand from the decay rates involved,
$a_1$ exchange will be much suppressed compared to $\rho$
exchange except at very low energies. Nevertheless, we include the 
$a_1$ contributions to $V_{\mu\nu}$.

The resonance Lagrangian \cite{egpr89} and corresponding diagrams 
are displayed in App.~\ref{app:reso}. The Lagrangian gives also rise
to double-$\rho$ exchange diagrams. One of them is needed for ensuring 
gauge invariance beyond $O(p^4)$ accuracy but we include all of them
for consistency. In terms of the couplings defined in 
App.~\ref{app:reso}, the final result takes the form:
\begin{eqnarray} 
v_1 &=& \frac{F_V G_V}{F_\pi^2 M_\rho^2}\left\{2 + 2 M_\rho^2 
D_\rho^{-1}[(P-q)^2] + t D_\rho^{-1}(t) + 
t M_\rho^2 D_\rho^{-1}(t) D_\rho^{-1}[(P-q)^2] \right\}\nn
&+& \frac{F_V^2}{2 F_\pi^2 M_\rho^2}\left\{ -1 - M_\rho^2 
D_\rho^{-1}[(P-q)^2] + (P-q)^2 D_\rho^{-1}[(P-q)^2]\right\}\nn
&+& \frac{F_A^2}{F_\pi^2 M_{a_1}^2}(M_{a_1}^2 - M_\pi^2 + t/2)
D_{a_1}^{-1}[(p_-+k)^2] \label{eq:v1}\\
v_2 &=& \frac{F_V G_V t}{F_\pi^2 M_\rho^2}\left\{- D_\rho^{-1}(t) - 
M_\rho^2 D_\rho^{-1}(t) D_\rho^{-1}[(P-q)^2] \right\}\nn
&+& \frac{F_V^2}{2 F_\pi^2 M_\rho^2}\left\{-1 - M_\rho^2 
D_\rho^{-1}[(P-q)^2] - (P-q)^2 D_\rho^{-1}[(P-q)^2]\right\}\nn
&+& \frac{F_A^2}{F_\pi^2 M_{a_1}^2}(M_{a_1}^2 - M_\pi^2 - p_-\cdot k)
D_{a_1}^{-1}[(p_-+k)^2] \label{eq:v2}\\ 
v_3 &=& \frac{F_A^2}{F_\pi^2 M_{a_1}^2}D_{a_1}^{-1}[(p_-+k)^2] 
\label{eq:v3}\\ 
v_4 &=& - \frac{2 F_V G_V}{F_\pi^2}D_\rho^{-1}(t) D_\rho^{-1}[(P-q)^2]
+ \frac{F_V^2}{F_\pi^2 M_\rho^2}D_\rho^{-1}[(P-q)^2] \label{eq:v4}~.
\end{eqnarray}  
The resonance propagators are given by 
\begin{eqnarray} 
D_\rho(t) &=& M_\rho^2 - t - i M_\rho \Gamma_\rho(t) \label{eq:prop}\\ 
D_{a_1}(t) &=& M_{a_1}^2 - t - i M_{a_1} \Gamma_{a_1}(t) \no
\end{eqnarray} 
in terms of the momentum dependent width \cite{spanff1}
\begin{equation}	 
\hspace*{-.2cm} \Gamma_\rho(t) = \frac{M_\rho t}{96 \pi F_\pi^2}\left[
(1 - 4 M_\pi^2/t)^{3/2}\theta(t-4 M_\pi^2)+\frac{1}{2}
(1 - 4 M_K^2/t)^{3/2} \theta(t-4 M_K^2)\right]~.
\end{equation} 
For $\Gamma_{a_1}(t)$ we use the parametrization of Ref.~\cite{ks90}.
Because of the already mentioned suppression of $a_1$ exchange neither
the form of $\Gamma_{a_1}(t)$ nor its on-shell value (we take
$\Gamma_{a_1}(M^2_{a_1})=$ 0.5 GeV) are of much importance in
practice. 

To define the decay amplitude (\ref{eq:T}) completely, we still have
to specify the pion form factor $f_+(t)$. We use the results
of Ref.~\cite{spanff1} where the chiral form factor of $O(p^4)$ was
matched to  the resonance region
\begin{equation}
f_+(t)=M_\rho^2 D_\rho^{-1}(t)
\exp{ \left[2{\tilde H}_{\pi\pi}(t)+{\tilde H}_{KK}(t)
\right]} ~. \label{eq:span}
\end{equation} 
The subtracted loop function ${\tilde H}_{PP}(t)$ is defined in 
App.~\ref{app:kin}.

The pion form factor (\ref{eq:span}) exhibits the correct low-energy
behaviour to $O(p^4)$ by construction \cite{spanff1}. To check the
low-energy limit of the vector amplitudes $v_1, v_2$ ($v_3, v_4$ arise
only at $O(p^6)$), one has to insert the resonance dominance relations
\cite{egpr89}
\begin{eqnarray}
L_9 &=& \frac{F_V G_V}{2 M_\rho^2} \label{eq:resdom} \\
L_{10} &=&  \frac{F_A^2}{4 M_{a_1}^2}-\frac{F_V^2}{4 M_\rho^2} 
\no ~.
\end{eqnarray}

\section{Decay rate and spectrum}
\label{sec:rates} 
\renewcommand{\theequation}{\arabic{section}.\arabic{equation}}
\setcounter{equation}{0}
The differential rate for the decay $\tau^- \to \nu_\tau 
\pi^- \pi^0 \gamma$ is given by
\begin{eqnarray} 
d\Gamma &=& \displaystyle\frac{1}{4 m_\tau (2\pi)^8}
{\displaystyle\sum_{\rm spins}^{}} |T|^2d_{\rm LIPS}
\label{eq:rate}\\
d_{\rm LIPS} &=& \frac{d^3q}{2E_\nu}\frac{d^3p_-}{2E_{\pi^-}}
\frac{d^3p_0}{2E_{\pi^0}}\frac{d^3k}{2k^0} \delta^{(4)}(P-q-p_--p_0-k)
~. \no
\end{eqnarray}  
The decay rate is dominated by bremsstrahlung of soft
photons. To be sensitive to the dynamics of the proper radiative
transition, a substantial cut on photon energies will be necessary. 

We distinguish in the following between the full amplitude presented
in the previous section and what we shall call ``complete 
bremsstrahlung'', i.e. the amplitude with 
\begin{eqnarray} 
v_1=v_2=v_3=v_4=a_1=a_2=0 ~. \label{eq:compbrems}
\end{eqnarray} 
Since we use the expression (\ref{eq:span}) for the pion form factor 
in both cases, the correct Low limit (\ref{eq:Low}) is guaranteed in 
both scenarios. 

In Table \ref{tab:rate}, we display
the decay rate for three different lower cuts on the photon energy.
\TABLE[ht]{
%  \centering
\caption{Branching ratios $BR(\tau^- \to \nu_\tau \pi^- 
\pi^0 \gamma)$ for different cuts $E_\gamma^{\rm min}$ on the
  photon energy. The second column corresponds to the full amplitude
  of Sec.~\protect\ref{sec:amp}, the third one to the complete
  bremsstrahlung approximation defined in the text.}
%  \vskip 0.2 in
\begin{tabular}{|c|c|c|}  \hline
$E_\gamma^{\rm min}$ (MeV) & \hspace*{.5cm} BR(full) 
\hspace*{.5cm} & \hspace*{.5cm} BR(brems) \hspace*{.5cm} \\
\hline
  100  &  $8.4 \cdot 10^{-4}$ & $8.0 \cdot 10^{-4}$ \\
  300  &  $1.8 \cdot 10^{-4}$ &  $1.5 \cdot 10^{-4}$\\
  500  &  $3.8 \cdot 10^{-5}$  &  $2.6 \cdot 10^{-5}$ \\
\hline
\end{tabular}
\label{tab:rate} }
In Fig.~\ref{fig:spectrum} we compare the $t$-distribution for
$E_\gamma \ge E_\gamma^{\rm min} =$ 300 MeV (in the $\tau$ rest frame) 
for the two options: the complete spectrum is shown as the full curve, 
the distribution corresponding to complete bremsstrahlung is given by 
the dashed curve. The interpretation of the spectrum is straightforward. 
The dominant peak is almost exclusively due to bremsstrahlung off the 
$\pi^-$. The secondary peak receives contributions from bremsstrahlung
off the $\tau$ lepton but for a sufficiently high cut on the photon
energy the spectrum is also sensitive to the additional resonance 
exchange contributions in $V_{\mu\nu}$ due to the diagrams 
in Fig.~\ref{fig:vdiag}. 
\FIGURE[ht]{
\setlength{\unitlength}{1cm} 
\centering
\begin{picture}(16,10)
\put(-0.5,0){\makebox(14,10)[lb]{\epsfig{figure=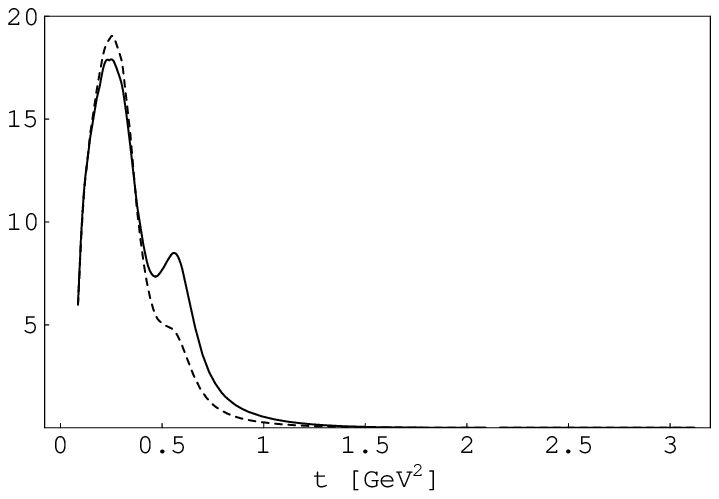,height=10cm}}}
\put(8.5,7){\Large{$E_\gamma \ge 300$ MeV}}
\end{picture}
\caption{Decay spectrum (arbitrary units) as function of
t, the invariant mass squared of the two pions, for
photon energies bigger than $E_\gamma^{\rm min}$=300 MeV (in the
$\tau$ rest frame): full amplitude (full curve) vs. complete
bremsstrahlung (dashed curve).}
\label{fig:spectrum} }

There are at present no published results for the radiative decay
under discussion. Table \ref{tab:rate} and Fig.~\ref{fig:spectrum}
suggest $E_\gamma^{\rm min} \sim$ 300 MeV as a reasonable value for
the cutoff in photon energy to test the nontrivial dynamics of the
radiative decay. For smaller values, rate and spectrum
become more and more dominated by (complete) bremsstrahlung. For
larger cutoffs the branching ratios become too small. 

We add a few comments on the theoretical status of the amplitude. 
Although our approach is well motivated from the low-energy structure
predicted by QCD we cannot claim more than what it is worth: a
realistic model for the intermediate-energy region that should be 
confronted with experiment. For instance, there is no guarantee that 
effects that only show up at $O(p^6)$ and higher in the low-energy 
expansion are fully accounted for. The situation is different for
complete bremsstrahlung. In this case, the amplitude is completely
determined by the pion form factor (\ref{eq:span}) which is known to
describe the data very well in the region of interest \cite{spanff1}.
Therefore, whenever complete bremsstrahlung dominates (for 
$E_\gamma^{\rm min} < $ 100 MeV), rate and spectrum can be predicted
in a model independent way.

\section{Radiative corrections for hadronic vacuum polarization}
\label{sec:vacpol} 
\renewcommand{\theequation}{\arabic{section}.\arabic{equation}}
\setcounter{equation}{0}

The experimental precision in $\tau$ decays has made very accurate
determinations of hadronic vacuum polarization possible. In
particular, a precision of 1 $\%$ has been achieved \cite{adh,dhprep} 
for the contribution of hadronic vacuum polarization to the anomalous
magnetic moment of the muon. This accuracy makes a careful
investigation of isospin violating and electromagnetic corrections
mandatory.

In a recent letter \cite{cen1}, we have calculated the electromagnetic
one-loop corrections for the two-pion decay of the $\tau$ lepton. Once
the infrared divergences are removed by adding the decay rate due to
soft photons, this loop correction is in principle well defined.
However, a sizable uncertainty remains in general because the
treatment of radiative events depends on the experimental setup. 
Without specifying those experimental conditions, we had to assign a
rather big uncertainty to the radiative corrections.
At best, the procedure of Ref.~\cite{cen1} may
be applied when only radiative events with photon energies below a 
sufficiently small energy cutoff are included. In this case, the
leading Low approximation to the decay rate, as contained for 
instance in the widely used Monte Carlo program PHOTOS \cite{was}, 
may be expected to provide a first approximation. But even in this
case, the specific value of the cutoff in photon energy would still
have to be implemented in the analysis of Ref.~\cite{cen1}.

For the precision needed in the determination of hadronic vacuum
polarization, a more careful analysis is mandatory. This is especially
true for the case where all radiative events are included in the 
two-pion data sample as for instance in the ALEPH experiment 
\cite{ALEPH}. When photons of all energies are considered the leading Low
approximation for the radiative rate is of course no longer justified.

For this purpose, we will use the model of Sec.~\ref{sec:amp} to
calculate the radiative differential decay rate. To make the
cancellation of infrared divergences transparent, we divide the
calculation into two parts. The infrared divergent decay rate in 
leading Low approximation will be calculated analytically. When added 
to the equally infrared divergent electromagnetic one-loop correction 
\cite{cen1} for the non-radiative decay, the infrared divergences
cancel and the limit of vanishing photon mass can be taken. The
second part of the radiative rate, containing in particular the
subleading Low terms of $O(k^{-1})$, will be calculated numerically.
This second part is infrared convergent and the photon mass can be set
to zero from the start.

In the notation of our letter \cite{cen1}, we set out to calculate the
electromagnetic correction function $G_{\rm EM}(t)$ defined by
\begin{equation} 
\label{eq:dGamma} 
\displaystyle\frac{d \Gamma
%(\tau^-\to \pi^0 \pi^- \nu_\tau[\gamma])
_{\pi \pi [\gamma]}}{dt} = 
\displaystyle\frac{G_F^2 m_\tau^3 S_{\rm EW}|V_{ud}|^2}{384 \pi^3}
 (1 - 4 M_\pi^2/t)^{3/2} \, (1-\frac{t}{m_\tau^2})^2  
(1+\frac{2t}{m_\tau^2}) |f_+(t)|^2  \, G_{\rm EM} (t) ~.
\end{equation} 
The decay rate here is the fully inclusive one, i.e., with photons of
all possible energies included. $S_{\rm EW}$ is a short-distance
electroweak correction factor \cite{MS88}. We also recall that we work 
in the isospin limit in this section.

In a first step, we consider only the leading Low approximation of 
$O(k^{-2})$ to the
differential decay rate for the radiative process:
\begin{eqnarray} 
d \Gamma &=& \displaystyle\frac{\alpha G_F^2 |V_{ud}|^2}
{32 \pi^7 m_\tau} |f_+(t)|^2 D(t,u) \times \label{eq:dGLow} \\
&&\hspace*{-1cm} \left\{\displaystyle\frac{2 P\cdot p_-}{(P\cdot k
-M_\gamma^2/2)(p_-\cdot k + M_\gamma^2/2)} -
\displaystyle\frac{m_\tau^2}{(P\cdot k - M_\gamma^2/2)^2} - 
\displaystyle\frac{M_\pi^2}{(p_-\cdot k + M_\gamma^2/2)^2}
\right\}d_{\rm LIPS} ~.\no
\end{eqnarray} 
The kinematics of the radiative decay and the function $D(t,u)$ are
given in App.~\ref{app:kin}. The Lorentz invariant phase space element
$d_{\rm LIPS}$ is defined in Eq.~(\ref{eq:rate}). Here and in the
following, we neglect terms that will not contribute to the final
result for $M_\gamma \to 0$.

Integration over neutrino and photon momenta leads to the three-fold
differential rate
\begin{eqnarray} 
d \Gamma &=& \displaystyle\frac{\alpha G_F^2 |V_{ud}|^2}
{64 \pi^4 m_\tau^3} |f_+(t)|^2 D(t,u) \times \label{eq:dGtux} \\
&& \left\{2 P\cdot p_- I_{11}(t,u,x)
- m_\tau^2 I_{20}(t,u,x) - M_\pi^2 I_{02}(t,u,x)
\right\}dt~du~dx \no
\end{eqnarray} 
with 
\begin{equation} 
I_{mn}(t,u,x) = \displaystyle\frac{1}{2\pi} \displaystyle\int 
\displaystyle\frac{d^3q}{2q^0}\displaystyle\frac{d^3k}{2k^0}
\displaystyle\frac{\delta^{(4)}(P-q-p_--p_0-k)}
{(P\cdot k - M_\gamma^2/2)^m(p_-\cdot k + M_\gamma^2/2)^n} ~.  
\end{equation} 

In the next step we perform the integration over $x$, the invariant
mass squared of photon and neutrino. Here we must distinguish between
two different regions in the $t-u$ plane. For $(t,u)$ in the Dalitz
plot of the non-radiative decay, the photon mass must be kept nonzero
and the lower limit for the $x$-integration is in fact
$M_\gamma^2$. For $(t,u)$ that cannot be accessed in the non-radiative
decay the lower limit is given by $x_-(t,u)$ in (\ref{eq:xpm}).
The resulting contribution to the rate is infrared finite. The latter
case can only occur for $t\le m_\tau^2 M_\pi/(m_\tau - M_\pi)=$ 0.27
GeV$^2$. The corresponding contribution to the spectrum is 
very small. We will therefore not display the explicit
formula for this part of the spectrum here but it will be
included in the correction function $G_{\rm EM}(t)$.
The upper limit of the $x$-integration is always given by 
$x_+(t,u)$ in (\ref{eq:xpm}).

The double differential decay rate in the leading Low approximation, 
for $(t,u)$ in the non-radiative Dalitz plot and
integrated over all photon momenta, takes the form
\begin{eqnarray} 
d \Gamma &=& \displaystyle\frac{\alpha G_F^2 |V_{ud}|^2}
{64 \pi^4 m_\tau^3} |f_+(t)|^2 D(t,u) \times \label{eq:dGtu1} \\
&& \left\{J_{11}(t,u,M_\gamma)+J_{20}(t,u,M_\gamma) +
J_{02}(t,u,M_\gamma)\right\}dt~du ~.\no
\end{eqnarray} 
Neglecting as always terms that vanish for $M_\gamma \to 0$, the
functions $J_{mn}$ are given by
\begin{eqnarray} 
J_{11}(t,u,M_\gamma) &=& \log(\frac{2 x_+ (t,u) \bar{\gamma}}{M_\gamma})
\displaystyle\frac{1}{\bar{\beta}}\log(\frac{1+\bar{\beta}}
{1-\bar{\beta}}) \nn
&+& \displaystyle\frac{1}{\bar{\beta}}\left\{ Li_2(1/Y_2) - Li_2(Y_1)+
\log^2(-1/Y_2)/4 - \log^2(-1/Y_1)/4 \right\}\nn
J_{20}(t,u,M_\gamma) &=& \log(\frac{M_\gamma (m_\tau^2 - t)}
{m_\tau x_+ (t,u)}) \nn
J_{02}(t,u,M_\gamma) &=& \log(\frac{M_\gamma (m_\tau^2 + M_\pi^2 -t-u)}
{M_\pi x_+ (t,u)}) \label{eq:Jmn}~.
\end{eqnarray}
The various quantities in these functions are defined in 
App.~\ref{app:kin}.

The remaining part of the radiative decay rate, containing all terms 
except the leading Low terms, is
calculated numerically. Putting everything together, the double
differential rate for the photon-inclusive two-pion decay is given by
\begin{equation} 
\label{eq:dGtu2} 
\displaystyle\frac{d \Gamma_
%(\tau^-\to \pi^0 \pi^- \nu_\tau[\gamma])
{\pi \pi [\gamma]}}{dt~du} = 
\displaystyle\frac{G_F^2 S_{\rm EW}|V_{ud}|^2}{64 \pi^3 m_\tau^3}
|f_+(t)|^2  D(t,u)\left\{1 + 2 f_{\rm loop}^{\rm
elm}(u,M_\gamma) + g_{\rm rad} (t,u,M_\gamma)\right\}.
\end{equation} 
The electromagnetic loop amplitude $f_{\rm loop}^{\rm elm}
(u,M_\gamma)$ was calculated in Ref.~\cite{cen1} and is reproduced in 
App.~\ref{app:kin}. The radiative part of the rate is determined by
the function $g_{\rm rad} (t,u,M_\gamma)$ with
\begin{eqnarray} 
g_{\rm rad} (t,u,M_\gamma) &=& g_{\rm brems} (t,u,M_\gamma) +
g_{\rm rest} (t,u) \label{eq:grad}\\
g_{\rm brems} (t,u,M_\gamma) &=& \displaystyle\frac{\alpha}{\pi}
[J_{11}(t,u,M_\gamma)+J_{20}(t,u,M_\gamma)+J_{02}(t,u,M_\gamma)] ~. 
\end{eqnarray} 
The (numerically calculated) function $g_{\rm rest} (t,u)$ accounts
for the infrared finite remainder of the rate, i.e., all terms except
the ones contained in $g_{\rm brems} (t,u,M_\gamma)$.
As it must be, the sum $2 f_{\rm loop}^{\rm elm} + g_{\rm brems}$ 
is independent of the photon mass and therefore infrared finite
as well.

To obtain the correction function $G_{\rm EM}(t)$, we still have to
integrate over $u$ \cite{cen1}:
\begin{equation} 
G_{\rm EM}(t) = \frac{ \displaystyle\int_{u_{\rm min}(t)}^{u_{\rm
 max}(t)} du \, D(t,u) \, \Delta(t,u) }{\displaystyle\int_{u_{\rm
 min}(t)}^{u_{\rm max}(t)} du \, D (t,u)} ~,
\label{eq:GEM}
\end{equation} 
with 
\begin{equation} 
\Delta(t,u) = 1 + 2 f_{\rm loop}^{\rm elm}(u,M_\gamma) + 
g_{\rm brems} (t,u,M_\gamma) + g_{\rm rest} (t,u)~.
\end{equation} 
The limits of integration $u_{\rm min,max}(t)$ can also be found in 
App.~\ref{app:kin}.

Since only the sum $2 f_{\rm loop}^{\rm elm} + g_{\rm brems}$ is
physically meaningful two distinct pieces can be distinguished in 
$G_{\rm EM}(t)$. Setting $g_{\rm rest}$ to zero, we obtain a
function $G_{\rm EM}^{(0)}(t)$ that describes radiative
corrections in the leading Low approximation. In 
Fig.~\ref{fig:GEM} we plot both the complete result $G_{\rm EM}(t)$ 
and the (leading) bremsstrahlung approximation $G_{\rm EM}^{(0)}(t)$. 
\FIGURE[ht]{
\setlength{\unitlength}{1cm} 
\centering
\begin{picture}(14,10)
\put(-1.2,0){\makebox(13,10)[lb]{\epsfig{figure=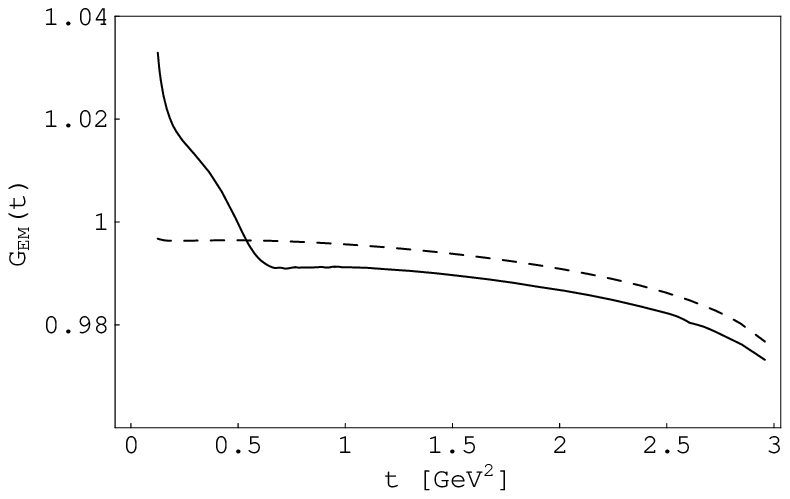,height=10cm}}}
\end{picture}
\caption{Correction function $G_{\rm EM}(t)$ defined in 
(\protect\ref{eq:dGamma}) and (\protect\ref{eq:GEM}). The full curve
corresponds to the complete radiative amplitude, the dashed curve
describes the leading Low approximation.}
\label{fig:GEM}
}

The obvious conclusion is that the non-leading terms in the Low
expansion cannot be neglected when considering the photon-inclusive
spectrum and rate. One should keep in mind that the function 
$G_{\rm EM}(t)$ multiplies the non-radiative spectrum in 
(\ref{eq:dGamma}) that is strongly peaked around $t=M_\rho^2\simeq$
0.6 GeV$^2$. 

Where does the difference between the
two curves in Fig.~\ref{fig:GEM} mainly come from? The answer is that 
it comes almost exclusively from the subleading Low terms in
(\ref{eq:Low}). In other words, the amplitudes 
$v_1, \dots, v_4, a_1, a_2$ have practically no impact on
$G_{\rm EM}(t)$. In fact, $G_{\rm EM}(t)$ for 
$$
v_1=v_2=v_3=v_4=a_1=a_2=0
$$
could not be distinguished from the full result in 
Fig.~\ref{fig:GEM}. This conclusion corresponds to the 
findings in Sec.~\ref{sec:rates} that the specific features of the 
model for the radiative transition only show up for a sufficiently 
large lower cut on photon energies.

In the present case where photons of all energies are included,
complete brems\-strahlung is an excellent approximation.
However, as Fig.~\ref{fig:GEM} documents very clearly, the leading 
Low approximation ($O(k^{-2})$ in rate) corresponding
to the function $G_{\rm EM}^{(0)}(t)$ is not a valid approximation.
The complete bremsstrahlung approximation 
($v_1=v_2=v_3= v_4=a_1=a_2=0$ in (\ref{eq:vmunu}), (\ref{eq:amunu})) 
is uniquely determined by the pion form factor $f_+(t)$. The correction
function $G_{\rm EM}(t)$ is therefore effectively independent of the 
details of the model of Sec.~\ref{sec:amp} for the radiative transition.

Finally, we emphasize once again that the analysis in this section
applies to the photon-inclusive case only. Moreover, even the leading
Low approximation $G_{\rm EM}^{(0)}(t)$ (dashed curve in
Fig.~\ref{fig:GEM}) cannot be compared directly with the corresponding
function in Ref.~\cite{cen1} that contained only the loop contribution
(with an ad hoc prescription for the infrared divergent part). In
contrast, the curves in Fig.~\ref{fig:GEM} refer to the complete
electromagnetic corrections, i.e., to the (infrared finite) sum of
loop and radiative contributions with photons of all energies
included.

\section{Isospin violating corrections for $\mbf{a_\mu}$}
\label{sec:amu} 
\renewcommand{\theequation}{\arabic{section}.\arabic{equation}}
\setcounter{equation}{0}

The leading hadronic contribution to the anomalous magnetic moment of
the muon $a_\mu=(g_\mu-2)/2$ is due to hadronic vacuum
polarization \cite{GdR69},
\begin{equation}
a_\mu^{\rm vacpol}=\displaystyle\frac{1}{4\pi^3}\displaystyle\int_{4
M_\pi^2}^{\infty} dt K(t) \sigma^0_{e^+ e^- \to  {\rm hadrons}}(t)~,
\label{eq:amuvp}
\end{equation} 
where $K(t)$ is a smooth kernel concentrated at low energies and
$\sigma^0_{e^+ e^- \to {\rm hadrons}}$ denotes the bare hadronic cross
section with QED corrections removed.  The low-energy structure of
hadronic vacuum polarization is especially important. In fact, about
70 $\%$ of $a_\mu^{\rm vacpol}$ is due to the two-pion intermediate
state for $4 M_\pi^2 \le t \le 0.8$ GeV$^2$ (see, e.g.,
Ref. \cite{narison}). A substantial improvement in the accuracy is
possible if one includes data from hadronic $\tau$ decays.
A precise link between hadronic spectral functions from $\tau$ decays 
and from the $e^+ e^-$ hadronic cross section requires the
calculation of radiative corrections as well as the inclusion of other
isospin breaking effects (both kinematical and dynamical in
origin). For the two-pion final state, we have parametrized 
\cite{cen1} the relation between the {\em bare} $e^+ e^-$ cross
section and the {\em observed} differential $\tau$ decay rate as
follows \footnote{The bare $e^+ e^-$ cross section directly provides
the two-pion contribution to the muon $g-2$ at $O(\alpha^2)$. 
We follow here the tacit convention that only explicit photonic
corrections are considered in the counting of powers of $\alpha$.
In other words, electromagnetic contributions to the charged pion mass
or to $\rho$-$\omega$ mixing are included in the lowest-order
contribution of $O(\alpha^2)$.}:
\beq
\sigma^{0}_{\pi \pi} = \left[ \frac{K_{\sigma} (t)}{K_{\Gamma} (t)} \, 
\frac{ d \Gamma_{\pi \pi [\gamma]}}{d t} \right] \times 
\frac{R_{\rm IB} (t) }{S_{\rm EW}}~,
\label{eq:sigma} 
\eeq
where  
\beqa
K_{\Gamma} (t) &=&   \frac{G_{F}^2 \, |V_{ud}|^2  \,  m_\tau^3}{ 384 \pi^3} \
\left( 1 - \frac{t}{m_\tau^2} \right)^2  \, 
\left( 1 + 2 \frac{t}{m_\tau^2} \right)  \nonumber \\
K_{\sigma} (t) &=&   \frac{\pi \, \alpha^2}{3 \, t} 
\eeqa
and the isospin breaking correction is in 
\beq
R_{\rm IB} (t) = \frac{1}{G_{\rm EM} (t)} \ \frac{ \beta_{\pi^+ \pi^-}^3}{ 
\beta_{\pi^+ \pi^0}^3} \ \left| \frac{ F_V (t)}{f_+ (t) } \right|^2 ~. 
\eeq
The factor $S_{\rm EW}=$ 1.0194 (taken at the scale $m_\tau$) 
takes into account the dominant short-distance
electroweak corrections \cite{MS88}.  $R_{\rm IB} (t)$ involves the
long-distance QED correction $G_{\rm EM}(t)$ (discussed in detail in
the previous section), the phase space correction factor
$\beta^3_{\pi^+ \pi^-}(t)/\beta^3_{\pi^0\pi^-}(t)$ (important in the
low-$t$ region), and the ratio of electromagnetic ($F_V$) over weak
($f_+$) pion form factors.  

In this section we discuss the isospin breaking
affecting the ratio of form factors.  Since our purpose is to relate the
observed weak decay to the bare two-pion cross section, we need to
include effects of order $(m_u - m_d)$ in both $F_V$ and $f_{+}$, while
we need explicit photonic corrections of order $\alpha$ only in
$f_{+}$. Keeping track of all relevant sources of isospin breaking at 
first order, Eq.~(\ref{eq:span}) generalizes to
\beqa
F_V(t)&=&
M_{\rho^0}^2 D_{\rho^0}^{-1}(t)  
\Bigg[
\exp{ 
\left(2{\tilde H}_{\pi^+ \pi^-}(t)+{\tilde H}_{K^+ K^-}(t)\right)} 
 - \frac{\theta_{\rho\omega}}{3 M_{\rho}^2}  \frac{t}{M_{\omega}^2 - 
t - i M_{\omega} \Gamma_{\omega} }
\Bigg] \qquad
\label{eq:span-zero}
\\
 f_+(t)&=&M_{\rho^+}^2 D_{\rho^+}^{-1}(t) \, 
% \Bigg[
\exp{ 
\left(2{\tilde H}_{\pi^+ \pi^0}(t)+{\tilde H}_{K^+ K^0}(t)
\right)} 
+ f_{\rm local}^{\rm elm} + \dots   \qquad \qquad 
% \Bigg]
\label{eq:span-plus}
\eeqa
%\end{minipage}
In Eq.~(\ref{eq:span-zero}) the main new ingredient compared to 
Ref.~\cite{cen1} is the inclusion of
the $\omega$ contribution to the electromagnetic form factor, following 
the notation of Ref.~\cite{Urech:1995ry}. This contribution
was not included in our previous work \cite{cen1} because it is
formally of higher order in the low-energy expansion. Here we include 
$\rho$-$\omega$ mixing together with all other numerically relevant 
isospin violating contributions. 
The mixing parameter can be extracted from the decay width for 
$\omega \to \pi^+ \pi^-$ leading to $\theta_{\rho\omega}= 
(- 3.9 \pm 0.3)\cdot 10^{-3}$ GeV$^2$ in the framework
of Ref.~\cite{Urech:1995ry}. A more direct determination comes from
the recent Novosibirsk data for $e^+ e^- \to \pi^+ \pi^-$ 
\cite{CMD2002}. Translating their fit for the mixing parameter into our 
notation gives  $\theta_{\rho\omega}= (- 3.3 \pm 0.5)\cdot 10^{-3}$ 
GeV$^2$ (the same value has been obtained in an independent fit of the
CMD-2 data \cite{Jorge}).
For the numerical discussion we take $\theta_{\rho\omega}= 
(- 3.5 \pm 0.7)\cdot 10^{-3}$ GeV$^2$. For the remaining parameters 
we use $M_{\omega} = 0.783$ GeV  and $\Gamma_{\omega} = 0.00844$ GeV.

Eq.~(\ref{eq:span-plus}) reproduces our previous result \cite{cen1} 
where $f_{\rm local}^{\rm elm}$ represents the structure dependent 
electromagnetic corrections to the low-$t$ part of the spectrum: 
\beqa
f_{\rm local}^{\rm elm} &=& \frac{\alpha}{4 \pi} \bigg[ - \frac{3}{2} 
- \frac{1}{2} \log \frac{m_{\tau}^2}{\mu^2} 
- \log \frac{M_{\pi}^2}{\mu^2} + 2 \log \frac{m_{\tau}^2}{M_{\rho}^2}
\nonumber \\
& & \qquad - (4 \pi)^2 
\left(- 2 K_{12}^{r} (\mu) + \frac{2}{3} X_1 
+ \frac{1}{2} {\tilde X}_6^{r} (\mu) \right) \bigg] \ .
\eeqa
$K_{12}$, $X_1$ and $X_{6}$ are LECs appearing in the 
effective Lagrangian of CHPT. 
Numerical estimates for the chiral couplings are reported in 
Ref.~\cite{cen1}.
The dots in Eq.~(\ref{eq:span-plus}) represent unknown structure
dependent corrections to the form factor, away from the threshold
region. At the moment there is no model independent method for
including such terms. However, our calculation correctly identifies 
the large UV logs (in $S_{\rm EW}$) and the IR logs (chiral logs in 
$f_{\rm local}^{\rm elm}$ and $f_{\rm loop}^{\rm elm}$), and 
therefore the neglected terms are expected to be numerically smaller.

In the numerical analysis we also account for isospin violating
effects due to mass and width differences between the charged and the
neutral $\rho$ mesons. The $\rho^+$-$\rho^0$ mass difference is unknown
but expected to be small \cite{bg96,ALEPH}. We follow Ref.~\cite{adh} and
assume $M_{\rho^+}-M_{\rho^0} = (0 \pm 1)$ MeV. In addition to the
two-pion contributions, there is also a possible width difference due
to radiative decays, dominated by the $\pi \pi \gamma$ 
modes. We include the measured $\pi \pi \gamma$ decay width of the 
$\rho^0$ of approximately 1.5 MeV \cite{pdg00} and assume a radiative
decay width difference $\Gamma(\rho^0 \to \pi^+ \pi^- \gamma)-
\Gamma(\rho^+ \to \pi^+ \pi^0 \gamma) = (0.45 \pm 0.45)$ MeV 
\cite{singer,hn95,adh}.
\FIGURE[ht]{
\setlength{\unitlength}{1cm} 
\centering
\begin{picture}(16,10)
\put(1.0,0){\makebox(14,10)[lb]{\epsfig{figure=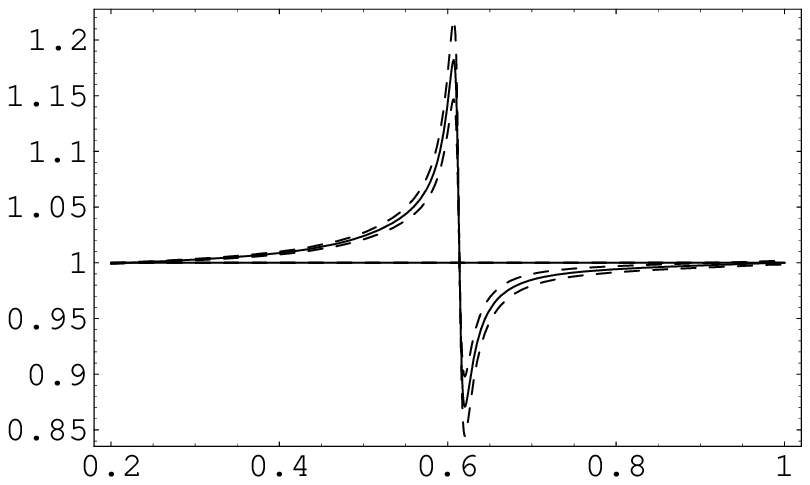,height=8.5cm}}}
\put(7.5,-0.5){\large{$t$ [GeV$^2$]}}
\put(-0.5,4.0){\large{ $|F_V / f_+|^2$ }}
\end{picture}
\vspace{0.2cm}
\caption{
The ratio $|F_V / f_+|^2$ for $\theta_{\rho\omega}= 
(- 3.5 \pm 0.7)\cdot 10^{-3}$ GeV$^2$. The full curve corresponds to
the mean value. For this plot we have assumed $M_{\rho^+}=M_{\rho^0}=
775$ MeV and we have neglected the radiative decay widths.} 
\label{fig:rff}
}

We can summarize our results graphically. In Fig.~\ref{fig:rff} we
plot the ratio $|F_V / f_+|^2 $ showing the sensitivity to 
$\theta_{\rho\omega}$. In Fig.~\ref{fig:rib} we plot the
full isospin violating correction function $R_{\rm IB}(t)$ containing 
the complete radiative correction $G_{\rm EM}(t)$, displaying also 
the small uncertainty due to the electromagnetic LECs.  

\FIGURE[ht]{
\setlength{\unitlength}{1cm} 
\centering
\begin{picture}(16,10)
\put(1.0,0){\makebox(14,10)[lb]{\epsfig{figure=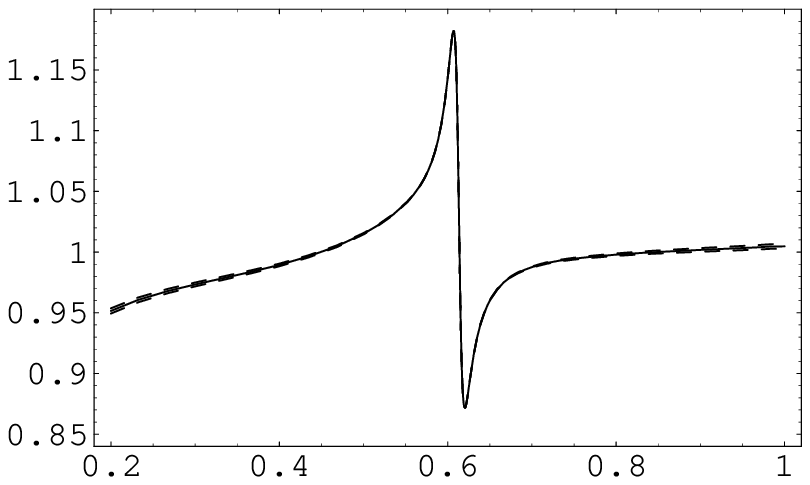,height=8.5cm}}}
\put(7.5,-0.5){\large{$t$ [GeV$^2$]}}
\put(0,4.0){\large{$R_{\rm IB}(t)$}}
\end{picture}
\vspace{0.5cm}
\caption{
Full correction factor $R_{\rm IB} (t)$ with
the complete radiative correction $G_{\rm EM}(t)$ (solid curve). 
The (almost indistinguishable) dashed curves correspond to 
variation of the chiral couplings within their error bars \cite{cen1}. 
In this plot we use  $\theta_{\rho\omega}= 
- 3.5 \cdot 10^{-3}$ GeV$^2$, $\rho$ masses 
and widths as in Fig.~\ref{fig:rff}. }
\label{fig:rib}
}

In order to investigate the impact of $R_{\rm IB}(t)$ on $a_\mu^{\rm
vacpol}$, we have evaluated for the two-pion contribution to
(\ref{eq:amuvp}) the difference between the correct expression 
for $\sigma^{0}_{\pi \pi}$ (full RHS of Eq.~(\ref{eq:sigma})) and the 
naive CVC expression (the term in square brackets on the RHS of
(\ref{eq:sigma})), up to a scale $t_{\rm max}$:
\beq
\Delta a_\mu^{\rm vacpol}=\displaystyle\frac{1}{4\pi^3}\displaystyle\int_{4
M_\pi^2}^{t_{\rm max}} dt K(t) \, 
\left[ \frac{K_{\sigma} (t)}{K_{\Gamma} (t)} \, 
\frac{ d \Gamma_{\pi \pi [\gamma]}}{d t} \right] \times 
\left(\frac{R_{\rm IB} (t) }{S_{\rm EW}} \, - \, 1 \right)~.
\label{eq:damu}
\eeq 
In Table~\ref{tab:shift}, we have analysed separately the effects 
of $S_{\rm EW}$ 
and of the three ingredients of $R_{IB}$: kinematics (KIN), 
radiative correction (EM) and form factor ratio (FF).
We find that the difference $\Delta a_{\mu}^{\rm vac pol}$ is quite
insensitive to $t_{\rm max}$ as long as  $t_{\rm
max} \ge 1$ GeV$^2$ (cf. Table~\ref{tab:shift}).  We can
summarize our results as follows:
\TABLE[ht]{
\caption{Contributions to $\Delta a_\mu^{\rm vacpol}$  from 
various sources of isospin violation (in units of $10^{-11}$) for 
different values of 
$t_{\max}$ (in units of GeV$^2$) as defined in Eq.~(\ref{eq:damu}).} 
\label{tab:shift}
\vspace{.5cm}
\begin{tabular}{|c|c|c|c|c|c|}
\hline 
 & & &  & & \\ 
\mbox{  } $t_{\rm max}$ \mbox{  } &  $S_{\rm EW}$  &  KIN    
& \mbox{ } EM \mbox{ }  &  FF & $ \Delta a_\mu^{\rm vacpol}$ (total)   \\ 
 & & &  & & \\
\hline
 & & &  & & \\
1 & - 95   &   - 75  & - 11 & 61 $\pm$ 26 $\pm$ 3
  & - 119  \\
2 & - 97 &   - 75  & - 10 &
 61 $\pm$ 26 $\pm$ 3   &  - 120 \\
3 & - 97  &   - 75  & - 10 &
 61 $\pm$ 26 $\pm$ 3   &  - 120\\
 & & &  & & \\
\hline 
\end{tabular}
}
\begin{itemize}
\item The short-distance correction $S_{\rm EW}=$ 1.0194 induces 
$\Delta a_\mu^{\rm vacpol}= - 97 \cdot 10^{-11}$.
\item The kinematical effect induces 
a shift $ \Delta a_\mu^{\rm vacpol}= - 75 \cdot 10^{-11}$.
\item The complete radiative corrections produce a shift
$ \Delta a_\mu^{\rm vacpol}= - 10 \cdot 10^{-11}$.  We emphasize once
more that this correction applies only for the case of a 
photon-inclusive two-pion data set as provided for instance by the
ALEPH experiment \cite{ALEPH}. If we had used the leading Low 
correction factor $G_{\rm EM}^{(0)}(t)$, the shift would be 
$\Delta a_\mu^{\rm vacpol}= + 16 \cdot 10^{-11}$ instead.
\item The ratio of form factors produces a positive shift
of approximately $61 \cdot 10^{-11}$.
Roughly 60 $\%$ of this effect is due to $\rho$-$\omega$ mixing, the
remainder comes from the $\rho^+$-$\rho^0$ width
difference (pion mass difference and radiative decay modes).
We add the errors due to the previously discussed uncertainties of 
$\rho$-$\omega$ mixing (8), $\rho^+$-$\rho^-$ mass difference (20) 
and radiative decay widths (14) in quadrature (the numbers in brackets 
denote the corresponding shifts in units of $10^{-11}$), but keep the 
small additional uncertainty in the electromagnetic LECs \cite{cen1} 
(3) separate. The final shift due to the form factor ratio is
therefore $ \Delta a_\mu^{\rm vacpol}= 
+ ( 61 \pm 26 \pm 3)  \cdot 10^{-11}$.  
\end{itemize}
Our results are based on the form factor representations of
Eqs.~(\ref{eq:span-zero}) and (\ref{eq:span-plus}).  Use of the
experimentally measured $f_{+}$, as well as a different treatment of
the form factor ratio \cite{adh}, produce slightly different
values for the individual contributions to $\Delta a_{\mu}^{\rm vac
pol}$. The overall picture emerging from the two treatments is
however quite consistent: there is a strong  cancellation of all 
effects included in $R_{\rm IB}(t)$. The results in
Table~\ref{tab:shift} put this statement on a quantitative basis. 

Collecting all isospin violating contributions, we obtain finally  
\begin{equation} 
\Delta a_\mu^{\rm vacpol}=  ( - 120  \pm 26 \pm 3) \cdot 10^{-11}~.
\label{eq:shift}
\end{equation}
We have not assigned an error for the
short-distance correction factor $S_{EW}$ because this is already
included in the error for the electromagnetic LECs. $S_{EW}$ is
in fact closely related to the LEC $X_6^r(M_\rho)$ after scaling
it down from $m_\tau$ to $M_\rho$ \cite{cen1,CKNRT02}. Altogether, we 
are confident that with the photon-inclusive two-pion decay of the 
$\tau$ as input the isospin breaking shift (\ref{eq:shift}) can be 
given to an accuracy of about $20\%$. The accuracy could be improved
considerably by better determinations of the $\rho$ mass 
and radiative decay width differences.

\section{Conclusions}
\label{sec:conc} 
\renewcommand{\theequation}{\arabic{section}.\arabic{equation}}
\setcounter{equation}{0}

Guided by the low-energy limit dictated by CHPT, we have 
constructed a meson resonance model which allows the investigation of the 
radiative $\tau$ decay $\tau^- \to \nu_\tau \pi^- \pi^0 \gamma$ for 
the whole range of momentum transfer to the hadronic system. We have shown 
that for a photon energy cut of approximately 300 MeV in the $\tau$
rest frame, the nontrivial 
part of the amplitude can be discriminated from the pure 
bremsstrahlung contribution in the decay spectrum.

As a second application of our model, we have considered the radiative 
corrections for the process $\tau^- \to \nu_\tau \pi^- \pi^0$.
This includes also real photon emission via the associated radiative 
decay. Such an investigation is relevant once  experimental data from 
hadronic $\tau$ decays are used for the determination of the hadronic 
vacuum polarization, which requires a reliable estimate of 
isospin violating effects. 

The contribution from real photon emission to the decay rate depends
on the experimental setup. If photons of all energies are 
included (as in the ALEPH data \cite{ALEPH}), the leading Low 
approximation is 
definitely not sufficient any more. On the other hand, complete 
bremsstrahlung turns out to be an excellent approximation in this
case. Compared to the naive CVC relation, we find a total isospin
violating shift of the
anomalous magnetic moment of the muon $a_\mu$  of 
$\Delta a_\mu^{\rm vacpol} = (- 120 \pm 26 \pm 3) \cdot 10^{-11}$.
The first error is due to the combined uncertainties of
$\rho$-$\omega$ mixing, $\rho^+$-$\rho^0$ mass and (radiative decay)
width differences. The comparatively small second error is based
on available estimates of the low-energy constants.

\acknowledgments 
We thank Michel Davier and Andreas H\"ocker for helpful discussions
and for information on the data analysis of the ALEPH experiment. We
are grateful to Jorge Portol\'{e}s for pointing out to us the new
determination of the $\rho$-$\omega$ mixing angle by the CMD-2
collaboration. The work of V.C. has been supported in part by MCYT, 
Spain (Grant No. FPA-2001-3031) and by ERDF funds from the European 
Commission.

\medskip\medskip
\newcounter{zaehler}
\renewcommand{\thesection}{\Alph{zaehler}}
\renewcommand{\theequation}{\Alph{zaehler}.\arabic{equation}}
\setcounter{zaehler}{1}
\setcounter{equation}{0}

%\newpage
\section{Resonance contributions}
\label{app:reso}
The resonance exchange contributions considered in Sec.~\ref{sec:amp} 
are derived from the standard chiral resonance Lagrangian
\cite{egpr89} for vector and axial-vector mesons (scalar exchange does 
not contribute to our decay):
\begin{eqnarray} 
\cL[V(1^{--}),A(1^{++})] &=& \cL_{\rm kin}[V,A]
\label{eq:resop4} \\
&+&  \dfrac{F_V}{2 \sqrt{2}}\langle V_{\mu\nu} f_+^{\mu\nu}\rangle + 
    \dfrac{iG_V}{\sqrt{2}} \langle V_{\mu\nu} u^\mu u^\nu\rangle 
 + \dfrac{F_A}{2 \sqrt{2}} \langle A_{\mu\nu} f_-^{\mu\nu} \rangle ~.
\no
\end{eqnarray} 
The Lagrangian is formulated in terms of $SU(3)$ tensor fields 
$V_{\mu\nu}, A_{\mu\nu}$ (not to be confused with the tensor
amplitudes of Sec.~\ref{sec:amp}). We refer to Ref.~\cite{egpr89} for
definitions and further details, in particular for the explicit form of
the gauge and chiral invariant kinetic terms that contain the
couplings bilinear in the resonance fields.

In addition to $M_\rho=0.775$ GeV, $M_{a_1}=1.23$ GeV and 
$\Gamma_{a_1}=0.5$ GeV we use for the resonance couplings the 
theoretically favoured values \cite{eglpr89}
\begin{eqnarray*} 
F_V=\sqrt{2} F_\pi = 0.13 ~{\rm GeV},\quad 
G_V=F_\pi/\sqrt{2} = 0.065 ~{\rm GeV},\quad
F_A=F_\pi = 0.0924 ~{\rm GeV}.
\end{eqnarray*} 
These values for $F_V, G_V$ are used in the energy dependent width 
(\ref{eq:prop}) leading to $\Gamma_\rho=\Gamma_\rho(M_\rho^2)=0.147$ 
GeV. 

The vector amplitudes $v_1, \dots, v_4$ are derived from the 
diagrams in Fig.~\ref{fig:vdiag}.
\FIGURE[ht]{ 
\centerline{\epsfig{file=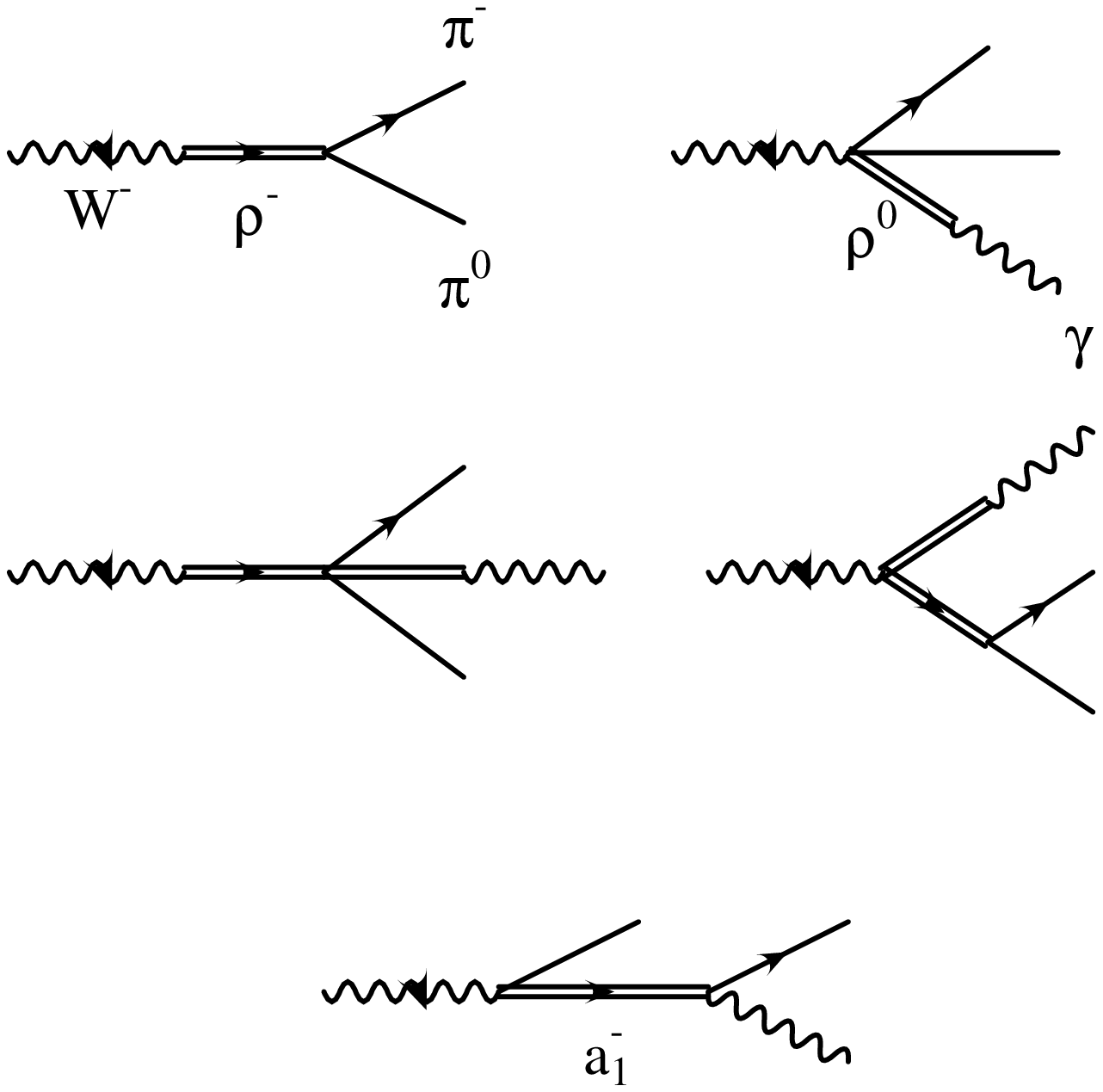,height=12cm}}
\caption{$\rho$ and $a_1$ exchange diagrams for the model of 
Sec.~\protect\ref{sec:amp}. In the first diagram, a photon is to be
appended wherever possible.}
\label{fig:vdiag}
}
 
\setcounter{equation}{0}
\addtocounter{zaehler}{1}
\section{Kinematics and loop corrections}
\label{app:kin}

\subsection{Kinematics}
After integration over neutrino and photon 4-momenta, 
the spin-averaged decay distribution for 
$$
\tau^-(P) \to \nu_\tau(q) \pi^-(p_1) \pi^0(p_2) \gamma(k)
$$
depends on three invariants\footnote{In the body of the text we use
the more suggestive notation $p_1=p_-$, $p_2=p_0$.}. We use 
\beq
t=(p_1 + p_2)^2~, \quad u=(P - p_1)^2~, \quad x=(q + k)^2 
\eeq
with $P^2=m_\tau^2$,  $p_i^2 = M_i^2$ ($i=1,2$).
In order to calculate differential rates and spectra, 
we need the physical region ${\cal D}$ in the form of a normal 
domain:
\beq
{\cal D} = \{ t_{\rm min}   \leq  t \leq   t_{\rm  max}, \  
u_{\rm min} (t)   \leq  u \leq   u_{\rm  max}(t), \ 
x_{\rm min} (t,u)   \leq  x \leq   x_{\rm  max}(t,u) \} \ , 
\eeq
where 
\beq
t_{\rm min}=  (M_1  + M_2)^2 ~,\qquad  t_{\rm  max}= m_{\tau}^2~.  
\eeq
In order to give the explicit form of $u_{\rm min/max} (t)$ and  
$x_{\rm min/max} (t,u)$ we introduce the auxiliary functions 
\beqa
u_{\pm} (t,x) & = & 
\frac{1}{2 \,  t} \Bigg[ 2 \, t \,  (m_{\tau}^2 + M_1^2) - (t + M_1^2 - M_2^2) 
 \, (m_{\tau}^2 + t - x) \nonumber \\
& \pm & \sqrt{ \lambda(t, M_1^2, M_2^2) \,  
\lambda(m_{\tau}^2, t, x) } \Bigg] 
\label{eq:upm} \\
x_{\pm} (t,u) & = & 
\frac{1}{2 \, M_1^2} \Bigg[ 2 \, M_1^2 \, (m_{\tau}^2 + t) - 
(t + M_1^2 - M_2^2) \, 
 (m_{\tau}^2 + M_1^2 - u) \nonumber \\ 
& \pm &  \sqrt{ \lambda(t, M_1^2, M_2^2) \,  
\lambda(u, M_1^2, m_{\tau}^2) } \Bigg] 
\label{eq:xpm}
\eeqa
with $\lambda(x,y,z) = x^2 + y^2 + z^2 - 2 \,  (x y + x z + y z )$.
In the following,  let us call ${\cal R}^{III}$ the region 
in the $t$-$u$ plane accessible in the non-radiative three-body decay, 
and ${\cal R}^{IV}$ the region accessible in the radiative decay. 
${\cal R}^{III}$ is given by
\beqa
\bar{u}_{\rm min} (t) &=& u_- (t,0)              \qquad     
 \qquad  \mbox{for} \quad    
 t_{\rm min}   \leq  t \leq   t_{\rm  max}   \\ 
\bar{u}_{\rm max} (t) &=& u_+ (t,0)              \qquad     
 \qquad  \mbox{for} \quad    
 t_{\rm min}   \leq  t \leq   t_{\rm  max}   
\eeqa 
while ${\cal R}^{IV}$ corresponds to  
\beqa
u_{\rm min} (t) &=& u_- (t,0)              \qquad         \qquad     
 \qquad \ \  \mbox{for} \quad    
 t_{\rm min}   \leq  t \leq   t_{\rm  max}  \\ 
 &  &   \nonumber \\
u_{\rm max} (t) &=& \left\{    
\begin{array}{ll} 
 u_+ (t,0)  &  \qquad  \quad  \mbox{for} \quad   
t_{*}   \leq  t \leq   t_{\rm  max}     
\\  (m_{\tau} - M_1)^2  &  \qquad    \quad  \mbox{for} \quad    
 t_{\rm min}   \leq  t \leq   t_{*} 
\end{array}
\right.   
\eeqa 
with 
\beq
t_{*} = \frac{m_{\tau} 
 ( m_{\tau} M_1 + M_2^2 - M_1^2 )}{m_{\tau} - M_1} \ .   
\eeq
Finally, for given $(t,u)$ the limits for $x$ are 
\beqa
x_{\rm max} (t,u) &=& x_+ (t,u)   \\ 
 &  &   \nonumber \\
x_{\rm min} (t,u) &=& \left\{    
\begin{array}{ll} 
 M_\gamma^2  &  \qquad  \mbox{for} \quad  (t,u) \in {\cal R}^{III}     
\\ x_- (t,u) &  \qquad  \mbox{for} \quad  (t,u) \in 
{\cal R}^{IV}\setminus{\cal R}^{III}~.      
\end{array}
\right. 
%\nonumber  
\eeqa 

The differential decay rate in Eq.~(\ref{eq:dGtu1}) involves 
the function 
\begin{equation}
D (t,u) = \frac{m_\tau^2}{2} (m_\tau^2 - t) + 2 M_1^2 M_2^2  
 - 2 \, u \, (m_\tau^2 - t + M_{1}^2  +  M_{2}^2 )  + 2 \, u^2 ~.
\end{equation}
Moreover, the following combinations of invariants appear in 
Eq.~(\ref{eq:Jmn}): 
\beq
Y_{1,2} = \frac{1 - 2 \bar{\alpha} \pm \sqrt{(1 - 2 \bar{\alpha} )^2 - (1 - 
\bar{\beta}^2 )} }{1  + \bar{\beta}}~,
\eeq
with 
\beqa
\bar{\alpha} &=& \displaystyle\frac{(m_\tau^2 - t)  
(m_\tau^2 + M_2^2 - t - u)}{
  (M_1^2 + m_\tau^2 - u)} \cdot \frac{\lambda (u, M_1^2, m_\tau^2)}
{2 \, \bar{\delta} }
\nonumber \\
  &  & \nonumber \\
\bar{\beta} &=& - \frac{\sqrt{ \lambda (u, M_1^2, m_\tau^2) 
   }  }{M_1^2 + m_\tau^2 - u   }  \nonumber \\
  &  & \nonumber \\
\bar{\gamma} &=& \frac{ \sqrt{  \lambda (u, M_1^2, m_\tau^2)    }}{
2 \, \sqrt{ \bar{\delta}} }     \nonumber  \\
  &  & \nonumber \\
\bar{\delta} &=& - M_2^4 m_\tau^2 + M_1^2 (m_\tau^2 - t) (M_2^2 - u) - 
t u ( - m_\tau^2 + t + u) \nonumber \\
& &  \qquad  + M_2^2 (- m_\tau^4 + t u + m_\tau^2 t 
+ m_\tau^2 u)  \nonumber 
\eeqa

\subsection{Loop functions}

The loop function $\tilde{H}_{PQ} (t,\mu)$ is given by 
\begin{eqnarray} 
\tilde{H}_{PQ} (t,\mu) &=& \frac{1}{F^2} {\cal R}e \ \Bigg[
 \frac{1}{12 t} \lambda (t,M_P^2,M_Q^2) \, 
\bar{J}^{PQ} (t) 
+ \frac{1}{18 (4 \pi)^2} (t - 3 \Sigma_{PQ}) \nn
&-&  \frac{1}{12} \left( \frac{2 \Sigma_{PQ} - t}{\Delta_{PQ}} (A_P(\mu) - 
A_Q(\mu))  - 2 (A_P(\mu) + A_Q(\mu)) \right) \Bigg] \ , 
\end{eqnarray} 
where
\begin{eqnarray}
\Sigma_{PQ} & = & M_P^2 + M_Q^2~, \qquad \Delta_{PQ} ~ = ~ M_P^2 -
M_Q^2  \nn 
A_P(\mu)  & = &  - \frac{M_P^2}{(4 \pi)^2} 
\log \frac{M_P^2}{\mu^2}  \nn
\bar{J}^{PQ} (t) & = & \frac{1}{32 \pi^2} \left[ 2 + \frac{\Delta_{PQ}}{t} 
\log \frac{M_Q^2}{M_P^2} - \frac{\Sigma_{PQ}}{\Delta_{PQ}} 
\log \frac{M_Q^2}{M_P^2}  \right. \nn
& -& \left.  \frac{\lambda^{1/2} (t,M_P^2,M_Q^2)}{t} \, 
\log \left( \frac{(t + \lambda^{1/2} (t,M_P^2,M_Q^2))^2 - \Delta_{PQ}^2}{(t - 
\lambda^{1/2} (t,M_P^2,M_Q^2))^2 - \Delta_{PQ}^2} \right) \right]~.
\end{eqnarray}

The one-loop virtual photon correction, contributing to the
differential spectrum (\ref{eq:dGtu2}), is given by
\footnote{ The last term in $f_{\rm loop}^{\rm elm}(u,M_\gamma) $ was
missing from Eq.~(16) of Ref.~\cite{cen1}. This typo did not affect the
numerical analysis of Ref.~\cite{cen1}. }
\beqa
f_{\rm loop}^{\rm elm}(u,M_\gamma) & = &  \frac{\alpha}{4 \pi} \bigg[
(u - M_{\pi}^2) \, {\cal A} (u) + 
(u - M_{\pi}^2 - m_{\tau}^2) \,  {\cal B} (u)  \nonumber \\*
&+ &   2 \, (M_{\pi}^2 + m_{\tau}^2 - u) \, {\cal C} (u,M_{\gamma}) 
+ 2 \, \log \frac{M_\pi m_\tau}{M_\gamma^2} 
\bigg]  \ .  \label{phloop} 
\eeqa
In terms of the variables 
\begin{equation}
 r_{\tau}  =  \frac{m_{\tau}^2}{M_{\pi}^2} \ , \ \ \ \
 y_{\tau} = 1 + r_{\tau} - \frac{u}{M_{\pi}^2} \ , \ \ \ \    
 x_{\tau} = \frac{1}{2 \sqrt{r_{\tau}}} (y_{\tau} - 
\sqrt{y_{\tau}^2 - 4 r_{\tau}}) \ ,     
\end{equation} 
and of the dilogarithm 
\begin{equation}
Li_2 (z) = - \int_{0}^{1} \frac{dt}{t} \log (1 - z  t)  \ , 
\end{equation}
the functions contributing to $f_{\rm loop}^{\rm elm} (u,M_\gamma)$ 
are given by
\begin{eqnarray}
{\cal A} (u)     &=&   \frac{1}{u} \left[ - \frac{1}{2} \log r_{\tau} + 
\frac{2 - y_{\tau}}{\sqrt{r_{\tau}}} \frac{x_{\tau}}{1 - 
x_{\tau}^2} \log x_{\tau} \right] \\
{\cal B} (u)     &=&   \frac{1}{u} \left[  \frac{1}{2} \log r_{\tau} + 
\frac{2 r_{\tau} - y_{\tau}}{\sqrt{r_{\tau}}} \frac{x_{\tau}}{1 -
 x_{\tau}^2} \log x_{\tau} \right] \\
{\cal C} (u,M_{\gamma})  &=&   \frac{1}{m_\tau M_\pi} \frac{x_{\tau}}{1 
- x_{\tau}^2}
\left[  - \frac{1}{2} \log^2 x_{\tau} + 2 \log x_{\tau} \log (1 - 
x_{\tau}^2) - 
\frac{\pi^2}{6} + \frac{1}{8} \log^2 r_\tau \right. \nn
 & +& \left.  Li_2 (x_{\tau}^2) + Li_2  (1 - 
\frac{x_{\tau}}{\sqrt{r_\tau}}) + 
Li_2 (1 - x_{\tau} \sqrt{r_\tau}) - \log x_{\tau} \log 
\frac{M_{\gamma}^2}{m_\tau M_\pi}
 \right]. \  \qquad 
\end{eqnarray}

\end{document}